\documentclass[aps,prb,twocolumn,superscriptaddress,showpacs]{revtex4}
\usepackage{epsfig}
\usepackage{amsmath}
\usepackage[dvipsnames,usenames]{color}

\begin{document}

\title{Low-energy electronic properties of Weyl semimetal quantum dot}

\author{Shu-feng Zhang}
\email[]{sps_zhangsf@ujn.edu.cn}
\affiliation{School of Physics and Technology, University of Jinan, Jinan, Shandong 250022, China}

\author{Chang-wen Zhang}
\affiliation{School of Physics and Technology, University of Jinan, Jinan, Shandong 250022, China}
\author{Pei-ji Wang}
\affiliation{School of Physics and Technology, University of Jinan, Jinan, Shandong 250022, China}

\author{Qing-Feng Sun}
\email[]{sunqf@pku.edu.cn}
\affiliation{International Center for Quantum Materials, School of Physics, Peking University, Beijing 100871, China}
\affiliation{Collaborative Innovation Center of Quantum Matter, Beijing 100871, China}

\begin{abstract}
It is necessary to study the properties of Weyl semimetal nanostructures for potential applications in nanoelectronics.
Here we study the Weyl semimetal quantum dot with a most simple model Hamiltonian with only two Weyl points. We focus on the low-energy electronic structure and show the correspondence to that of three-dimensional Weyl semimetal, such as Weyl point and Fermi arc.
We find that there exist both surface and bulk states near Fermi level.
The direct gap of bulk states reaches the minimum with the location determined by Weyl point.
There exists a quantum number with only several values supporting surface states,
which is the projection of Fermi arc.
The property of surface state is studied in detail, including circular persistent current,
orbital magnetic moment, and chiral spin polarization.
Surface states will be broken by a strong magnetic field and evolve into Landau levels gradually.
Simple expressions are derived to describe the energy spectra and electronic properties of surface states both in the presence and absence of magnetic field.
In addition, this study may help design a method to verify Weyl semimetal by separating out the signal of surface states since quantum dot has the largest surface-to-volume ratio.
\end{abstract}


\maketitle

\section{Introduction}

Topological materials have been one of the frontiers of condensed matter physics over the past decade,
with the focus shifting from topological insulators~\cite{Hasan2010RMP,Qi2011RMP} to topological semimetals,
one of which is Weyl semimetal proposed in 2011.~\cite{Wan2011PRB}
Its conduction and valence bands touch at some crystal momentums, the Weyl points,
in the first Brillouin zone near the Fermi level.
The low-energy excitations behave like Weyl fermions described by Weyl equation,~\cite{Weyl1929}
suggesting a platform to study and verify the properties of Weyl fermions.
Due to the ``no-go" theorem,~\cite{Nielsen1981NPB} Weyl points emerge in pairs in the crystal system,
being the singular points of Berry curvature with opposite topological charge or chirality.
There will be Fermi-arc surface states terminated at the projection of Weyl points due to the bulk-boundary correspondence. Recently, both Weyl points and Fermi arc have been discovered by ARPES measurements in TaAs~\cite{Xu2015Science,Lv2015PRX} soon after the prediction,~\cite{Weng2015PRX,Huang2015Natcommun} and then in some other materials.~\cite{Xu2015nphys,Liu2015nmat,Xu2015CPL,Deng2016nphys,Belopolski2016ncomms,Bruno2016PRB,Wu2016PRB}

Plenty of attentions have been paid to Weyl semimetal,~\cite{Burkov2016nmater,Jia2016nmater} including
predicting new materials (especially those with less Weyl points locating near Fermi level or magnetic Weyl semimetal),~\cite{Xu2011PRL,Burkov2011PRL, Weng2015PRX,Huang2015Natcommun}
verifying novel behaviors of Weyl fermions,~\cite{Zhang2016ncomm,Hou1} tuning these unique properties for real applications by magnetic or optical methods.~\cite{Parameswaran2014PRX,Potter2014ncomm,Chen2015PRL,Chan2016PRL,Jiang2015PRL}
However, we notice that:
(1) The measurement of magnetic Weyl semimetal is still challenging.
All Weyl semimetal materials verified directly are time-reversal invariant,
~\cite{Xu2015Science,Lv2015PRX,Xu2015nphys,Liu2015nmat,Xu2015CPL,Deng2016nphys,Belopolski2016ncomms,Bruno2016PRB,Wu2016PRB} though magnetic Weyl semimetal is predicted earlier.~\cite{Wan2011PRB}
Due to the formation of magnetic domains, the direct ARPES measurement is still unavailable in the magnetic Weyl semimetal.
It is still in need to design an effective method to verify the magnetic Weyl semimetal.
The observation of Fermi arc provides a method to verify the Weyl physics, which may be realised in a nanostructure
with high surface-to-volume ratio.
(2) So far, only a few reports pay attention to the low-dimensional system of Weyl semimetal.~\cite{Igarashi2017PRB,Liu2017PRB,Bovenzi2017arXiv}
Due to the high mobility,\cite{highmob1,highmob2,highmob3,highmob4,highmob5}
dissipationless surface channels and exotic chirality anomaly,
Weyl semimetal may play an important role in nanoelectronics and other fields.
The study of Weyl semimetal nanostructures, including the electronic and transport properties,
may be helpful and necessary for promoting the potential applications.
And it's also a wonder what's the correspondence of those exotic behaviors of three-dimensional (3D) Weyl semimetal in low-dimensional systems and whether new phenomena would emerge.

Quantum dot (QD) is a zero-dimensional nanostructure with the largest surface-to-volume ratio.~\cite{Smith2010ACR,addGPG}
It has been used widely with mature technology for conventional semiconductors.
We believe that the study of Weyl semimetal QD will help reveal the evolution of properties of
3D Weyl semimetal in low-dimensional system,
provide a new method of verifying Weyl semimetal by separating signal of surface states especially in transport measurements and promote the potential applications in nanoelectronics.
As far as we know rare works focus on Weyl semimetal QD.
At the same time we notice that there have been lots of works on the topological insulator QD,~\cite{Chang2011PRL,Kundu2011PRB,Li2014PRB,Zhang2013MPL,Ferreira2013PRL,
Paudel2013PRB,Korkusinski2014srep,Potasz2015nanoLett,Xin2015PRB,Dyke2016PRB,Qi2016China}
focusing on the electronic structure,~\cite{Chang2011PRL,Li2014PRB,Kundu2011PRB}
orbital magnetic moment,~\cite{Potasz2015nanoLett}
transport behavior,~\cite{Zhang2013MPL,Xin2015PRB,Dyke2016PRB} applications in quantum computing,~\cite{Ferreira2013PRL} and so on.
Topological insulator QD has been fabricated successfully in experiments.~\cite{cho2012nanoLett,Jia2015nanoRes}
These works on topological insulator QD not only offer valuable lessons on the study of
Weyl semimetal QD but also suggest that Weyl semimetal QD will be a fruitful field and deserves more attentions.

The first step in the study of Weyl semimetal QD is to understand the electronic structure and property.
Generally, the confinement in the spatial distribution of electrons result in the discrete quantization of energy levels and Coulomb interaction if there is several electrons in the QD.
In this paper we focus on the first effect only.
For simplicity, but still universal, we have chosen the model with only two Weyl points located on the $z$-axis which breaks the time reversal symmetry.
The geometry is chosen to be a cylinder to simplify the numerical calculation without losing the universality for properties being protected topologically.
The confinement in the $z$ (axial) direction and rotation invariance around the axial direction result in good quantum numbers $n_z$ and $j_z$, which are the projections of quantized momentum and total angular momentum along $z$-direction, respectively.
The direct energy gap of bulk states meets the minimum at certain $n_z$ and $j_z$,
which is determined by the location of Weyl point.
There will be both surface and bulk states near Fermi level.
The surface state distributes on the side surface but absent on the top and bottom surfaces of the QD.
It emerges only for several values of $n_z$ which are determined
by the spacing between the two Weyl points.
The surface states are similar with Fermi-arc surface states and
can be taken as its correspondence in QD system.
We have analyzed the properties of the surface state in detail.
The band of surface states depends linearly on $j_z$,
leading to chiral persistent currents and orbital magnetic moments.
Its local spin direction is approximately parallel to the current indicating a correspondence of spin-momentum locking in 3D case.
Besides we also study the size effect.
There will be no surface state in an ultra thin Weyl semimetal QD, which suggests that it's possible to control surface states by tuning its thickness.
Magnetic field will break the surface state and lead to Landau quantization gradually.
Energy bands in these two regimes, the surface state regime and Landau level regime, are fitted well with the derived expressions approximately.
The location of the crossover of the two regimes is also fitted which provides a criterion to determine
whether the magnetic filed or the system size is large enough to realize Landau quantization.

The rest of the paper is organized as follows.
In Sec. II, we give the model Hamiltonian and describes the method to solve the electron structure of
the cylinder Weyl semimetal QD.
In Sec. III, we show the numerical results and analyses in the absence of the magnetic field.
The effect of the magnetic field is discussed in Sec. IV.
Finally, we provide a brief discussion and conclusion in Sec. V.

\section{The Model Hamiltonian and Formalism }

\subsection{The Model Hamiltonian}

In this paper, we study a model of Weyl semimetal with only two Weyl points, protected by inversion symmetry while with time reversal symmetry broken.
The Hamiltonian in momentum space can be written as~\cite{Yang2011PRB,Lu2015PRB}
\begin{equation}
H=\tilde{\Delta}_{z}\sigma_{z}+A ( k_{x}\sigma_{x}+ k_{y}\sigma_{y} ),
\label{EqHc}%
\end{equation}
in which the Zeeman term $\tilde{\Delta}_{z}$ is given as
\begin{equation}
\tilde{\Delta}_{z} = M - \frac{m_0}{2} (k_{x}^{2}+k_{y}^{2}+k_{z}^{2}).
\end{equation}
$\sigma_{x,y,z}$ are the spin Pauli matrices. $A$, $M$ and $m_0$ are model parameters.
There are two energy bands which are obtained as
\begin{equation}
E_{\pm}(\mathbf{k})=\pm\sqrt{\tilde{\Delta}_{z}^{2}+A^{2}(k_{x}^{2}+k_{y}^{2})}.%
\end{equation}
There is an energy gap if $M/m_{0}<0$, and it lies in a normal insulator phase.
However, if $M/m_{0}>0$, these two bands will touch at two isolated points on $z$-axis,
${\bf k}_w = (0,0,\pm k_0)^T$, with $k_{0} = \sqrt{2M/m_{0}}$.
This two-fold degenerate point is named Weyl point since low-energy excitations near ${\bf k}_w $
can be described by Weyl equation approximately,~\cite{Weyl1929} which is derived to be
$  H_{\pm}=\mp m_{0}k_{0}\delta{k}_{z}\sigma_{z}+A(\sigma_{x}\delta k_{x}+\sigma_{y}\delta k_{y}) $ in our model, with the new wave vector $(\delta k_x, \delta k_y, \delta k_z) = {\bf k} - {\bf k}_w$ measured from the corresponding Weyl point.
Weyl point is the singular point of Berry curvature in momentum space and has a topological charge determined
by the flux of Berry curvature, $\mp\mathrm{sign} (m_{0})$ at $(0,0,\pm k_{0})^T$,~\cite{Lu2015PRB} which also describes the chirality or helicity of the excitations.
Weyl points always appear in pairs of opposite topological charge in bulk lattice.~\cite{Nielsen1981NPB}
The two Weyl points in our model are related by spatial inversion.~\cite{Yang2011PRB}

It is obvious that the Hamiltonian in Eq.(1) is invariant with rotation around the axial direction, as is also presented clearly in the energy dispersion since it depends on $k_x^2+k_y^2$ and $k_z^2$.
Besides, we find that there is also another antiunitary symmetry described by
\begin{eqnarray}\label{EqP}
  P = \sigma_x K,
\end{eqnarray}
where $K$ is the complex conjugate operator. By replacing ${\bf k}$ with $-i\partial_{\bf x}$, it is straightforward to prove that $P$ anticommutes with the Hamiltonian in Eq.~(\ref{EqHc}), which means $\{P,H\} =0$ in coordinate representation.
This symmetry leads to a constrain on the energy band, $E_-({\bf k}) = -E_+(-{\bf k})$, consistent with the energy band given above.

Hamiltonians with the opposite Zeeman term are related by time reversal transformation $T$,
which indicates $H(-M,-m_0) = T H(M,m_0)T^{-1}$.
On the other hand, Hamiltonians with the opposite $A$ are related by rotating the spin space,
which means $H(-A)=\sigma_zH(A)\sigma_z$.
Therefore, we take both $A$ and $m_0$ to be real and consider both positive and negative $M$ cases.

\subsection{Formalism}

In this paper, we consider a cylinder QD with height $L$ and radius $R$.
We simulate this system by applying an infinite potential $V=\infty$ outside the dot region.
Therefore, it will be convenient to solve the Schr\"odinger equation in the cylindrical coordinate system.
The boundary condition becomes $\Psi(r,\varphi,0) = \Psi(r,\varphi,L) = \Psi(R,\varphi,z) = 0$
and $\Psi(r,\varphi,z)= \Psi(r,\varphi+2\pi,z)$,
with $r$, $\varphi$ and $z$ being the coordinates in the radial, tangential and axial direction, respectively.

In the absence of the translation invariance, momentum is not a good quantum number any more and
it should be replaced with $\hat {\bf k} = -i\partial_{\bf x}$ in the Hamiltonian in Eq.(1).
However, it will be enlightening to consider the case with only the confining potential along $z$-direction.
We will get a two-dimensional film with $k_{x,y}$ being good quantum numbers.
Then, we arrive at the problem of standard one-dimensional infinite potential well for fixed $k_{x}$ and $k_y$, which has standing wave eigenvectors denoted by $ k_z = \frac{n_{z}\pi}{L}$ with $n_z=1,2,3...$.
It indicates that the $z$-component can be separated.
Besides, the projection of total angular momentum along $z$-axis, $j_z$, is also a good quantum number
since the Hamiltonian and geometry are rotation invariant around the $z$-axis.
And these two good quantum numbers $n_z$ and $j_z$ ensure that variables $z$, $\varphi$ and $r$
can be separated from each other.
A careful analysis suggests that the eigenvector can be written as
\begin{subequations}
\begin{eqnarray}
&\Psi_{n_{z},j_z}(r,\varphi, z)= \phi_{n_{z}}(z) \psi^{(n_z)}_{j_z}(r,\varphi), \label{Eq_phia}\\
&\phi_{n_{z}}(z)=\sqrt{\frac{2}{L}}\sin(\frac{n_{z}\pi}{L}z), \label{Eq_phib}\\
&\psi^{(n_z)}_{j_z}(r,\varphi)= \frac{1}{\sqrt{2\pi}}e^{-i\frac{\sigma_z}{2}\varphi} e^{i j_z\varphi}
\left(\begin{array}{c}
  A^{(n_z ,j_z)}(r)\\B^{(n_z, j_z)}(r)
\end{array}\right) .
\end{eqnarray}\label{Eq170118_01}
\end{subequations}
The z-component of total angular momentum is $\hat j_{z} = \hat L_z + \hat s_z$, with $\hat L_z = -i\partial_\varphi$ and $\hat s_z=\sigma_z/2$ being the orbital and spin operators respectively.~\cite{note01}
The eigenvector of $\hat L_z $ is $e^{i m \varphi}/\sqrt{2\pi}$ with eigenvalue $m$.
It's straightforward to verify that $\hat j_{z} \psi^{(n_z)}_{j_z}(r,\varphi) = j_z \psi^{(n_z)}_{j_z}(r,\varphi)$.
For a more intuitive understanding, we rewrite the eigenvector as
\begin{align}
\psi^{(n_z)}_{j_z}(r,\varphi) = \left(
\begin{array}[c]{c}%
A^{(n_z)}_{m-1}(r)\frac{1}{\sqrt{2\pi}}e^{i(m-1)\varphi}\\
B^{(n_z)}_{m}(r)\frac{1}{\sqrt{2\pi}}e^{im\varphi}%
\end{array}
\right).\label{Eq170117_01}
\end{align}
For the spin-up component, $L_z=m-1$, $s_z=1/2$, while for the spin-down component $L_z=m$, $s_z=-1/2$.
Therefore, for the both components, we have the total angular momentum $j_z$ related to the parameter $m$ by
\begin{eqnarray}
j_z=m- \frac{1}{2}.
\end{eqnarray}

The Hamiltonian in cylindrical coordinate, $H(r,\varphi,z)$,
can be obtained by substituting the momentum operator in the form of ~\cite{Li2014PRB}
\begin{align}
k_{+}  &  =e^{i\varphi}(-i\partial_{r}+\frac{1}{r}\partial_{\varphi
}),\nonumber\\
k_{-}  &  =e^{-i\varphi}(-i\partial_{r}-\frac{1}{r}\partial_{\varphi
}),\nonumber\\
k_{x}^{2}+k_{y}^{2}  &  =-(\frac{1}{r^{2}}\partial^{2}_{\varphi}+\frac{1}%
{r}\partial_{r}+\partial_{r^{2}}).
\end{align}
With the wave function of fixed quantum numbers $n_z$ and $j_z$, Eqs.~(\ref{Eq170118_01}) and (\ref{Eq170117_01}) substituted, the Schr\"odinger equation can be reduced with only the radial component left. Then we have
the radial Schr\"odinger equation in the form of
\begin{align}\label{Eq161101_03}
H_{n_z,j_z}(r)\left(
\begin{array}
[c]{c}%
A_{m-1}(r)\\
B_{m}(r)
\end{array}
\right)  =E\left(
\begin{array}
[c]{c}%
A_{m-1}(r)\\
B_{m}(r)
\end{array}
\right).
\end{align}%
The quantum number $n_z$ in the eigenvectors $A_{m-1}^{(n_z)}(r)$ and $B_{m}^{(n_z)}(r)$ is not marked explicitly
in the above equation and it will also be the case in the following unless it causes confusion.
Then replacing $k_z$ with $n_z\pi/L$ and taking a unitary transformation $H = U H(r,\varphi, z)U^\dag $ with
the unitary matrix $U = e^{i\frac{\sigma_z}{2}\varphi} e^{-i j_z\varphi}$,
the radial Hamiltonian in Eq.(\ref{Eq161101_03}) can be obtained as
\begin{align}\label{eq10}
H_{n_z,j_z}(r)=\left(
\begin{array}
[c]{cc}%
M_{n_{z}}+\frac{m_{0}}{2}\hat J^{(m-1)}(r)
& -iA  \hat J^{(m)}_-(r)\\
iA \hat J^{(m-1)}_+(r)
& -[M_{n_{z}}+\frac{m_{0}}{2} \hat J^{(m)}(r)]
\end{array}
\right)
\end{align}
in which we have denoted the effective mass term
\begin{eqnarray}
   M_{n_z} = M - \frac{1}{2}m_{0} (\frac{n_z\pi}{L})^2
\end{eqnarray}
and several new operators
\begin{align}
 & \hat J^{(m)}(r) = -\frac{m^{2}}{r^{2}}+\frac{1}{r}\partial_{r}+\partial_{r^{2}},\nonumber\\
 & \hat J^{(m)}_-(r) = \partial_{r}+\frac{m}{r},\nonumber\\
 & \hat J^{(m)}_+(r) =-\partial_{r}+\frac{m}{r}.
\end{align}
The axial and angular components of the eigenvector given in Eqs.~(\ref{Eq_phib}) and (\ref{Eq170117_01}) satisfy the boundary and normalization conditions. For the radial component, these two conditions demand
\begin{eqnarray}\label{eq13}
&A_{m-1}(R) = B_m(R) =0, \nonumber\\
 & \int_0^R [|A_{m-1}(r)|^2 + |B_m(r)|^2] r dr =1.
\end{eqnarray}

The radial Schr\"odinger equation, Eq.(\ref{Eq161101_03}) can be solved
by expanding the wave function with an appropriate basis.
We notice that the first kind Bessel function of $m$-th order, $J_m(r)$, is the eigenvector of the operator $\hat J^{(m)}(r)$, while $\hat J^{(m)}_+(r)$ and $\hat J^{(m)}_-(r)$ are the raising and lowering operators of $J_m(r)$ respectively, which means
\begin{align}
\hat{J}^{(m)}(r)J_{m}(r) & =-J_{m}(r),\nonumber\\
\hat{J}^{(m)}_+(r)J_{m}(r)  &  =J_{m+1}(r),\nonumber\\
\hat{J}^{(m)}_-(r)J_{m}(r)  &  = J_{m-1}(r).
\end{align}
It suggests that an appropriate basis to expand the radial eigenvector can be constructed with the first kind Bessel function as~\cite{Chang2011PRL,Kundu2011PRB,Li2014PRB}
\begin{align}
\tilde J_n^{(m)}(r)  &  = \frac{J_{m}(x_{n}^{(m)}\frac{r}{R}) } {N_{n}^{(m)}},
\end{align}
where $x_n^{(m)}$ is the $n$-th node of the first kind Bessel function of $m$-th order, i.e. $J_m(x_n^{(m)})=0$.
And $N_{n}^{(m)}=\frac{RJ_{m+1}(x_{n}^{(m)})}{\sqrt{2}}$ is the normalization factor.
The orthonormal constraint is satisfied $\int_0^R \tilde J_n^{(m)}(r) \tilde J_{n^\prime}^{(m)}(r) rdr = \delta_{n,n^\prime}$.
Then we can expand the radial eigenvector as
\begin{align}
A_{m-1}(r)  &  =\sum_{n}A_{n}^{(m-1)}\tilde J_{n}^{(m-1)}(r),\nonumber\\
B_{m}(r)  &  =\sum_{n}B_{n}^{(m)}\tilde J_{n}^{(m)}(r),
\label{Eq170117_02}
\end{align}
where $A_{n}^{(m-1)}$ and $B_{n}^{(m)}$ are expansion coefficients satisfying the normalization
condition $ \sum_n |A_n^{(m-1)}|^2 + \sum_n|B_n^{(m)}|^2 = 1$.
The boundary condition of Eq.(\ref{eq13}) is satisfied automatically by using the basis $\tilde J_n^{(m)}(r)$.
The coefficients $A_{n}^{(m-1)}$ and $B_{n}^{(m)}$
can be obtained by substituting the expanded eigenvector Eq.~(\ref{Eq170117_02}) into the radial Schr\"odinger equation, Eq.~(\ref{Eq161101_03}), and then solving the eigenvalue problem. In the new basis of Bessel functions, the radial Schr\"odinger equation becomes
\begin{align}
\sum_{n^{\prime}}&[\langle\tilde J_{n}^{(m-1)}|H_{11}|\tilde J_{n^{\prime}}%
^{(m-1)}\rangle A_{n^{\prime}}^{(m-1)}+ \langle\tilde J_{n}^{(m-1)}|H_{12}|\tilde J_{n^{\prime}}^{(m)}\rangle B_{n^{\prime}}^{(m)}]
\nonumber\\ &=EA_{n}^{(m-1)},\nonumber\\
\sum_{n^{\prime}}&[\langle\tilde J_{n}^{(m)}|H_{21}|\tilde J_{n^{\prime}}^{(m-1)}\rangle
A_{n^{\prime}}^{(m-1)}+ \langle\tilde J_{n}^{(m)}|H_{22} |\tilde J_{n^{\prime}}^{(m)}\rangle B_{n^{\prime}}^{(m)}] \nonumber\\
&=EB_{n}^{(m)},
\label{Eq161101_01}
\end{align}
where $H_{ij}$ is the $(i,j)$ element of the Hamiltonian operator matrix $H_{n_z,j_z}(r)$
and we have used the Dirac bracket to represent the wave function for convenience.
In this discrete representation, each element of Hamiltonian matrix can be obtained as
\begin{align}\label{Eq161101_02}
&\langle\tilde J_{n}^{(m-1)}|H_{11}|\tilde J_{n^{\prime}}^{(m-1)}\rangle =[M_{n_{z}}-\frac{m_{0}}{2}(x_{n}^{(m-1)}/R)^{2}]\delta_{nn^{\prime}}, %
\nonumber\\
 & \langle\tilde J_{n}^{(m)}|H_{22}|\tilde J_{n^{\prime}}^{(m)}\rangle =-[M_{n_{z}}-\frac{m_{0}}{2}(x_{n}^{(m)}/R)^{2}]\delta_{nn^{\prime}}, \nonumber\\
&\langle\tilde J_{n}^{(m)}|H_{21}|\tilde J_{n^{\prime}}^{(m-1)}\rangle =\langle\tilde J_{n^{\prime}}^{(m-1)}|H_{12}|\tilde J_{n}^{(m)}\rangle^* \nonumber\\
& =iA \frac{x_{n^{\prime}}^{(m-1)}}{R} \frac{\int_{0}^{R}J_{m}(x_{n}^{(m)}r/R)J_{m}(x_{n^{\prime}}^{(m-1)}r/R)rdr}{N_{n}^{(m)}N_{n^{\prime}}^{(m-1)}} .
\end{align}
The number of basis functions used in the expansion is chosen to ensure the convergence of the energy levels near the Fermi level. It is adequate to take 800 basis functions in our calculation.

Finally, we point out that the phase of the eigenvector can be thus chosen, $A_{m-1}(r)$ is real
while $B_m(r)$ is imaginary.
This point can be seen clearly by rotating the spin space with a unitary matrix
\begin{eqnarray}\label{EqS}
  S = \left(\begin{array}[c]{cc}
  1& \\ &i
  \end{array}\right).
\end{eqnarray}
After this transformation, we can obtain a real radial Hamiltonian
$ H_{n_z,j_z}^\prime(r) = S H_{n_z,j_z}(r) S^\dag$ and therefore a real eigenvector
$(A^\prime_{m-1}, B^\prime_{m})^T = S (A_{m-1}, B_{m})^T$.
Then, we can set $A_{m-1}(r)$ real while $B_m(r)$ imaginary.

\section{Electronic structure and properties of Weyl semimetal QD}

This section gives the numerical results and analyses in the absence of external magnetic field.
Firstly, we study the energy spectra as a function of the total angular momentum $j_z$.
There will be a linear band in the energy gap regime if the system lies in the Weyl semimetal phase.
It is verified to be the band of surface states by plotting the electron density distribution and
calculating the dependence of energy on QD size.
Then, we further calculate the corresponding current density distribution, orbital magnetic moment and
the spin orientation to show the property of these surface states.
Finally, we show the projection of Fermi arc and Weyl point of 3D Weyl semimetal in the QD system.
Besides, we study the size effect as well.
In the numerical calculations, unless otherwise stated,
we take the cylinder QD with height $L=100$ nm and radius $R = 75 $ nm,
and the systemic parameters are set
$A=500$ ${\rm meV nm}$, $m_0 = 1000$ ${\rm meV nm^2}$, $M = 50 $ ${\rm meV}$ and $ -50$ ${\rm meV}$
for the Weyl semimetal phase and the normal insulator phase, respectively.

\subsection{Energy spectra}

\begin{figure}[ptb]
\centering
\includegraphics[width=1\columnwidth]{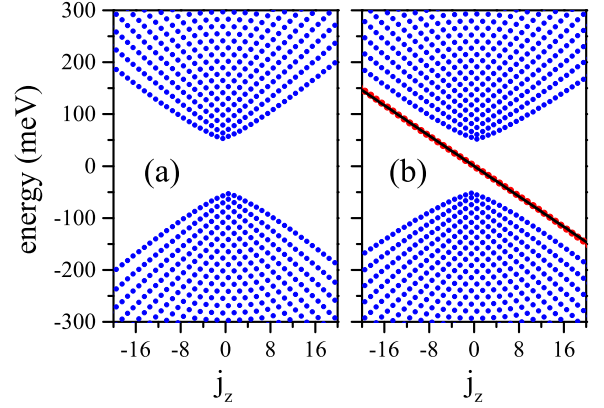}\newline
\caption{(Color online) The energy spectra as a function of angular momentum $j_z$ in the normal insulator phase ($M<0$) (a) and Weyl semimetal phase ($M>0$) (b) for sub-bands corresponding to $n_z=1$.
In panel (b), red dots show the linear band inside the gap regime
while the black line is a fitting line by Eq.~(\ref{EqEe}).}%
\label{figN01}%
\end{figure}

The confinement in axial, tangential and radial direction leads to energy quantization denoted by $n_z$, $j_z$
and $n_r$, respectively.
At first we consider the $n_z=1$ subband.
In Fig.~\ref{figN01}, we plot the energy spectra as a function of the total angular momentum $j_z$.
Fig.~\ref{figN01}(a) and Fig.~\ref{figN01}(b) correspond to the normal insulator and Weyl semimetal cases respectively.
It's clear that the energy spectra is not symmetric about $j_z$, which means $E_{n_z,-j_z}\neq E_{n_z,j_z}$.
However, with a detailed analysis we find that there exists an alternative relation written as
\begin{align}\label{Eq170413_01}
  E_{n_z, -j_z} = - E_{n_z, j_z}.
\end{align}
We have pointed out that the Hamiltonian in Eq.~(\ref{EqHc}) satisfies a certain symmetry determined by an antiunitary transformation $P = \sigma_x K$. For the radial Hamiltonian $H_{n_z,j_z}$ as shown in Eq.(\ref{eq10}), it results in
\begin{align}
  P H_{n_z,j_z}(r) P^\dag = - H_{n_z,-j_z}(r),
\end{align}
which leads to the constrain on energy spectra shown in Eq.~(\ref{Eq170413_01}) and constrain on corresponding eigenstates,
\begin{align}
  \psi_{-j_z}^{(n_z)}(r)   =\sigma_{x}[\psi_{j_z}^{(n_z)}(r)]^{*}.
\end{align}
Considering this relation, we are able to focus only on eigenstates with the positive angular momentum.

Another and the most significant feature is that the energy spectra is gapped for the normal insulator QD,
shown in Fig.~\ref{figN01}(a),
while there is a linear band (red dots) emerging even in the gap regime for the Weyl semimetal QD,
shown in Fig.~\ref{figN01}(b).
As is well known, the energy band due to edge or surface states of
topological insulators is a linear Dirac cone.~\cite{Hasan2010RMP,Qi2011RMP}
Therefore, it is natural to expect that the linear band here may correspond to surface states as well.
To test this point, we plot the distribution of electrons in this linear band
and study the dependence of these energy levels on QD radius $R$.

\begin{figure}[ptb]
\centering
\includegraphics[width=1\columnwidth]{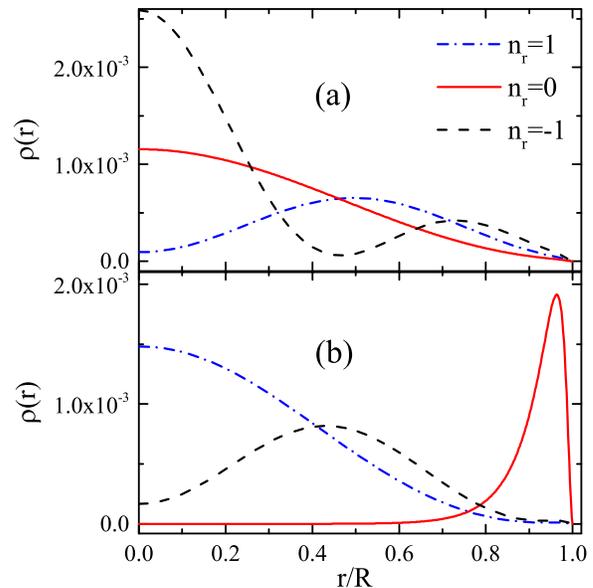}\newline
\caption{
(Color online) The radial electron density distributions of eigenstates
with $n_r=1$, $0$, and $-1$ for the normal insulator phase (a) and Weyl semimetal phase (b).
For the normal insulator phase, the three bands are the first conduction band ($n_r=1$) and
the first two valence bands ($n_r=0,-1$).
For the Weyl semimetal phase, the bands with $n_r=1$, $0$, and $-1$ are the first conduction band,
linear band, and the first valence band, respectively.
The other quantum numbers of the eigenstates are $j_z=\frac{1}{2}$ and $n_z=1$.}%
\label{figWave01}%
\end{figure}

For an electron in the eigenstate described by Eq.~(\ref{Eq170118_01}),
the density distribution $\rho(r,\varphi,z) = \Psi^\dag(r,\varphi,z) \Psi(r,\varphi,z)$ is given as,
\begin{eqnarray}
  \rho(r,\varphi,z) = \frac{2{\rm sin}^2(n_z\pi z/L)}{L} [ |A_{m-1}^2(r)| + |B_m^2(r)| ].
\end{eqnarray}
It is independent of $\varphi$ because of the rotation invariance around the axial direction.
The distribution in the axial direction is described by a sinusoidal function.
Then the unknown is only the radial distribution which is determined numerically and plotted in Fig.~\ref{figWave01} (b) for an electron in the linear band ($n_r=0$) and two neighbouring bands ($n_r=\pm1$)
with angular momentum $j_z=1/2$.
It's obvious that the red line, eigenvector from the linear band, locates near the side surface of the QD,
which is consistent with the feature of surface states.
But the other eigenvectors spread over the whole QD region, which is the feature of bulk states.
Fig.~\ref{figWave01} (a) plots the corresponding electron density distributions of the first conduction band ($n_r=1$) and the first two valence bands ($n_r=0$, $-1$) of the normal insulator QD
with $j_z=\frac{1}{2}$. All three states spread over the whole QD displaying bulk state behavior as expected.

\begin{figure}[ptb]
\centering
\includegraphics[width=1\columnwidth]{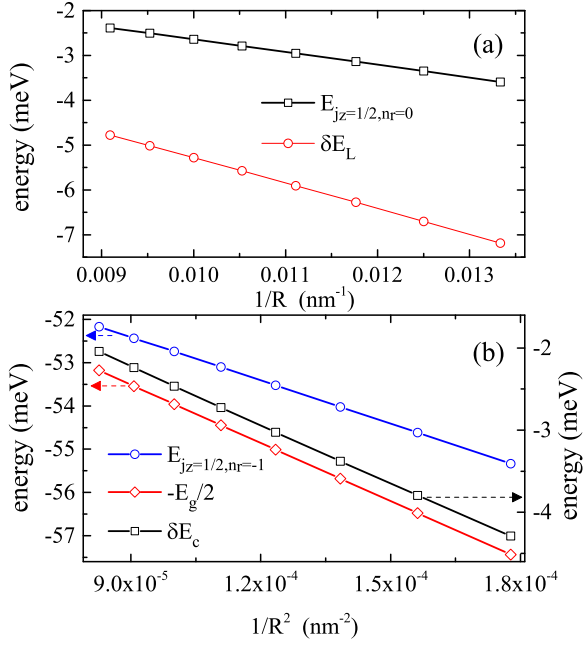}\newline
\caption{(Color online) The dependence of energy levels $E_{j_z,n_r}$,
energy spacing ($\delta E_L$ or $\delta E_c$), and direct gap $E_g$
on QD radius $R$ for the surface states (a) and bulk states (b).
}
\label{figN02}%
\end{figure}

The dependence of the energy levels on the radius of Weyl semimetal QD is calculated and shown in Fig.~\ref{figN02}.
Fig.~\ref{figN02}(a) focuses on the linear band ($n_r=0$).
The energy level $E_{j_z=1/2,n_r=0}$ and energy spacing $\delta E_L = E_{j_z=1/2,n_r=0} - E_{j_z=3/2,n_r=0}$ depend linearly on the inverse of the radius, $1/R$,
as expected for surface states on the side surface.
However, the direct gap $E_g = E_{j_z=1/2,n_r=1}-E_{j_z=1/2,n_r=-1}$, the energy level $E_{j_z=1/2,n_r=-1}$ and spacing $\delta E_c = E_{j_z=1/2,n_r=-1}-E_{j_z=3/2,n_r=-1}$ of the first conduction band depend linearly on the inverse of the cross-sectional area $1/R^2$, as shown in Fig.~\ref{figN02}(b), which is the bulk state behavior.~\cite{Chang2011PRL}

From the electron density distribution and energy dependence on QD radius,
it concludes that the linear band corresponds to the surface state.
It emerges since the topology of the Weyl semimetal QD and the vacuum is different just like
the surface state of TIs.~\cite{Hasan2010RMP,Qi2011RMP}
However, this surface state distributes only on the side but not on both ends of the cylinder QD.
It can be easily understood since the projection of two Weyl points on both ends will coincide
and cancel with each other due to the opposite topological charge.~\cite{Burkov2016nmater}
Besides, the surface state is quantized here and fortunately the liner dispersion remains.

It is able to derive an analytical expression to describe the band of surface states
by supposing a trial wave function~\cite{Chang2011PRL} in the form of
\begin{equation}\label{Eq170523_01}
\psi(r)= \frac{1}{\sqrt{2}}\left(
\begin{array}
[c]{c}%
1\\
-i \cdot sign(A)
\end{array}
\right) \tilde \psi(r) .
\end{equation}
The phase is chosen according to numerical results and the unitary transformation $S$ given in Eq.~(\ref{EqS}).
We assume that it is a perfect surface state locates at $r=R_L=c_LR$, which means
\begin{eqnarray}\label{Eq170523_02}
  \tilde \psi^*(r) \tilde\psi(r) \approx \frac{\delta(r-R_L)}{R_L}.
\end{eqnarray}
Substituting this trial function into the radial Schr\"odinger equation, Eq.~(\ref{Eq161101_03}),
then we can derive an approximate expression of the linear band
\begin{eqnarray}\label{EqEe}
  E=- j_z\frac{|A|}{\tilde c_{L}R}   + j_z \frac{m_{0}}{2\tilde c_{L}^{2}R^{2}}.
\end{eqnarray}
We have replaced the coefficient $c_L$ with $\tilde c_L$ manually to phenomenally describe
the effect not included in the trial wave function, such as the extension of wave function in the QD
and the unequal weight of spin components.
It's clear that it depends linearly on angular momentum $j_z$.
The second term is proportional to $1/R^2$ but with a much little weight factor,
therefore it depends linearly on QD radius $R$ approximately.
With the second term, Eq.~(\ref{EqEe}) also describes the interplay of bulk and surface characteristics.
The fitting line due to Eq.~(\ref{EqEe}) is plotted in Fig.~\ref{figN01}(b) (see the black solid line).
We can see that a good fitting is realized with coefficient $\tilde c_L \approx 0.9$.
However, we determine that $c_L\approx0.96$ or $0.93$ from the maximum and median of the electron density plotted in Fig.~\ref{figWave01}(b), respectively.
It indicates that the extension of wave function decreases the effective radius of the surface
state just as expected.

\subsection{Properties of the surface state}

In this subsection, we study the electronic properties of the surface state.
We focus on the current density distribution, orbital magnetic moment and the spin polarization.
The surface state with $j_z=\frac{1}{2}$ is taken as an example.

\subsubsection{Current and magnetic moment}

Here we show that there is a persistent vortex current due to the surface state.
The local probability current density
$\vec{J}(\vec{r})$ can be derived via the equation of continuity %
\begin{equation}
\frac{\partial\rho}{\partial{t}}+\nabla\cdot\vec{J}=0,
\end{equation}
in which $\rho(\vec{r})=\Psi^{\dag}(r,\varphi, z)\Psi(r,\varphi, z)$ is the probability density of electrons.
Here $\vec{r}=(x,y,z)=(r\sin\varphi, r\cos\varphi,z)$ is the 3D vector.
With a straightforward process, we arrive at
\begin{equation}
\vec{J}=\frac{-2}{\hbar}\mathrm{Im}[\frac{m_{0}}{2}\Psi^{\dag}\sigma
_{z}\nabla\Psi]+\frac{A}{\hbar}\mathrm{Re}[\Psi^{\dag}(\vec{e}_{x}\sigma
_{x}+\vec{e}_{y}\sigma_{y})\Psi].
\end{equation}
Due to the cylindrical geometry and rotation invariance,
only the tangential current $J_\varphi$ does not vanish and it is independent of $\varphi$.
For the eigenvector given in Eqs.~(\ref{Eq170118_01}) and (\ref{Eq170117_01}),
the tangential current $J_\varphi(r,z) $ becomes
\begin{eqnarray}
J_{\varphi}&=\left[-\frac{m_{0}}{2\pi\hbar}\psi_{j_z,n_r}^{(n_z) \dag}(r)\left(
\begin{array}
[c]{cc}%
m-1 & 0\\
0 & -m
\end{array}
\right)  \frac{1}{ r}\psi_{j_z,n_r}^{(n_z)}(r) \right. \nonumber\\
&\left. +\frac{A}{2\pi\hbar}\psi_{j_z,n_r}^{(n_z)\dag }(r) \sigma_y\psi_{j_z,n_r}^{(n_z)}(r)\right] \frac{2{\rm sin}^2(n_z\pi z/L)}{L} .
\label{Eq170109_03}%
\end{eqnarray}
The eigenvector will vanish on the axis of the cylinder QD, $\psi_{j_z,n_r}^{(n_z)}(r=0)=0$, except for the spin-up (spin-down) component of the state with $j_z=\frac{1}{2}$ ($j_z=-\frac{1}{2}$). Then it is evident from Eq.~(\ref{Eq170109_03}) that the current density $J_{\varphi}$ on the axis vanishes, which is consistent with the physical intuition.

\begin{figure}[ptb]
\centering
\includegraphics[width=1\columnwidth]{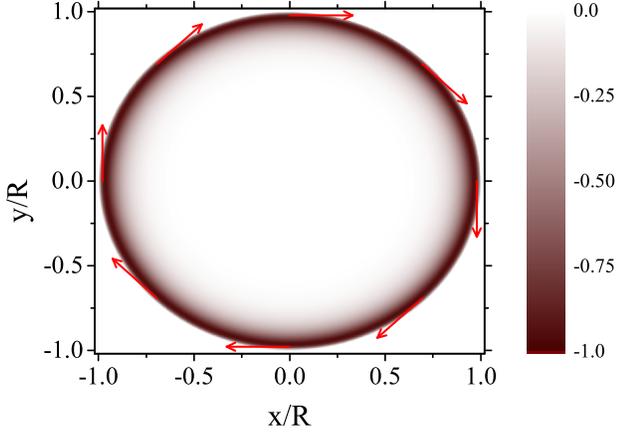}
\newline
\caption{(Color
online) The probability current density distribution of the surface state
$n_r=0$, $j_z=\frac{1}{2}$, $n_{z}=1$ in the horizontal plane.
Red arrows show directions of both current and spin orientation.  }%
\label{figN03}%
\end{figure}

The distribution of $J_{\varphi}(r,z)$ in the axial direction is described by a sinusoidal function explicitly as seen in Eq.~(\ref{Eq170109_03}). Therefore we calculate and plot only the current density distribution in the horizonal plane
in Fig.~\ref{figN03}.
The current locates around the side which is consistent with the density distribution of surface state.
We find that there will always be $J_\varphi(r,z)<0$ in the whole QD region
which corresponds to a clockwise probability current and an anticlockwise electric current.

The probability current $I_\varphi$ can be calculated by integrating over the cross-section
\begin{eqnarray}\label{EqIint}
   I_\varphi = \int_0^R\int_0^L  J_\varphi(r,z) dr dz.
\end{eqnarray}
And then we have the electric current $I_e = -eI_\varphi$
with $e$ being the magnitude of the electron charge.
On the other hand with the Hellmann-Feynman theorem, we can calculate the derivative of eigen-energy
with respect to angular momentum $j_z$. We derive that these two quantities are related by~\cite{note01}
\begin{equation}
I_{e} = -\frac{e}{h}\frac{dE_{n_{z}j_zn_r}}{dj_z},
\label{Eqc}
\end{equation}
in which $h=2\pi\hbar$ is the Plank constant.
Similar expressions also apply for the current of the Bloch state and Andereev bound state
in a Josephson junction.
It indicates that the directions of currents due to all surface states are identical regardless of
positive or negative angular momentums $j_z$ because of the negative slope of the linear
surface band shown in Fig.~\ref{figN01}(b).
In fact the flow direction is determined by the nontrivial topology of the system.
For the two-dimensional Weyl semimetal film with $n_z$ fixed, it is a Chern insulator,~\cite{Burkov2011PRL}
and its Chern number can be derived to be $C = ( \rm{sign}(M_{n_z}) + \rm{sign}(m_0) ) /2$.~\cite{Lu2010PRB}
So there will be a clockwise, zero, and anticlockwise current on the side surface for a positive,
zero, and negative Chern number.~\cite{book_Shen}

\begin{figure}
\centering
\includegraphics[width=1\columnwidth]{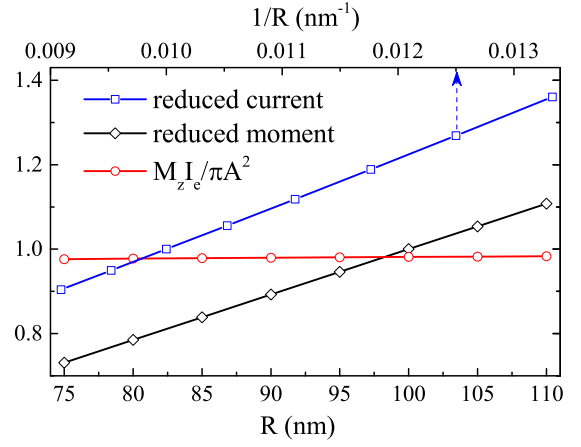}\\
\caption{(Color online) The reduced current, magnetic moment and their product as a function of
QD radius $R$ due to the surface state. The current and magnetic moment are reduced
by the value at $R=100$ nm, $\frac{I_e}{I_e(R=100nm)}$ and $\frac{M_z}{M_z(R=100nm)}$, respectively.
$\frac{M_z I_e}{\pi A^2}$ is in unit of $\frac{e^2}{h^2}$.
The surface state is the same as that in Fig.~\ref{figN03}. }
  \label{figM}
\end{figure}

A circular current results in an orbital magnetic moment defined as
$\vec{M} = \frac{1}{2}\int\vec{r} \times (-e\vec{J}) d\vec{r}$.
Here the moment due to the vortex current aligns in the axial direction,
and can be calculated by
\begin{align}
M_z =  -e \pi\int_{0}^{R}\int_0^L r^{2}J_{\varphi}(r,z) dr dz.
\label{Eq170304_03}
\end{align}
The numerical results are plotted in Fig.~\ref{figM} to show the evolution with QD radius $R$.
It suggests that the magnetic moment depends linearly on the QD radius
which is consistent with the result of the topological insulator QD.\cite{Potasz2015nanoLett}
The linear dependence of magnetic moment on $R$ is the same as that of a massless Dirac fermion
in a quantum ring.~\cite{Zarenia2010PRB}
But the case is different for a Schr\"odinger particle in the ring,
where the moment is independent of the ring size.~\cite{Potasz2015nanoLett}
It indicates that the surface state in Weyl semimetal QD can be seen as Weyl fermions confined in the side surface.
On the other hand, we find that the current depends linearly on the inverse of QD radius $1/R$
as shown in Fig.~\ref{figM}.
It suggests that the product of current and magnetic moment may be a constant.
The numerical result in Fig.~\ref{figM} indicates that
\begin{eqnarray}\label{EqIM}
  \frac{M_z I_{e}}{\pi A^2} \approx \frac{e^2}{h^2}.
\end{eqnarray}

By supposing a perfect surface state, we can derive an analytical current formula
and explain the dependence of current and magnetic moment on QD radius approximately.
Here the perfect surface state means that the radial component of the wave function located
at a certain $R_L$ and it can be written as
\begin{eqnarray}\label{Eq170411_01}
\psi(r)=\left(
\begin{array}
[c]{c}%
A_{L}\\
B_{L}%
\end{array}
\right) \tilde\psi(r),
\end{eqnarray}
where $\tilde\psi(r)$ is given in Eq.~(\ref{Eq170523_02}).
This wave function is similar with that defined in Eq.~(\ref{Eq170523_01})
but with no constrain on the weight on two spin components.
The normalization condition is realized by setting $|A_{L}|^{2}+|B_{L}|^{2}=1$.
Substituting the trial wave function given in Eq.~(\ref{Eq170411_01}) into the current formula in Eqs.~(\ref{Eq170109_03}) and (\ref{EqIint}), we derive
\begin{equation}\label{EqI}
hI_{\varphi}=-\frac{|A|}{R_{L}}\sqrt{1-4\langle s_{z}\rangle^{2}}+\frac{m_{0}}{2R_{L}^{2}}-\frac{2m_{0}}{R_{L}^{2}}j_z\langle s_{z}\rangle ,%
\end{equation}
where $\langle s_{z}\rangle =(|A_{L}^{2}|-|B_{L}^{2}|) / 2$ is the mean value of $z$-component of spin. The numerical calculation suggests that $\langle s_{z}\rangle $ is small, shown in Fig.\ref{figs} at $B=0$, which indicates that the current is nearly independent of angular momentum $j_z$. With $\langle s_z \rangle$ neglected, the current formula becomes
\begin{equation}\label{EqI2}
hI_{\varphi}=-\frac{|A|}{R_{L}} + \frac{m_{0}}{2R_{L}^{2}}.
\end{equation}
The first term is dominate and it depends on QD radius linearly,
which explains the numerical results in Fig.~\ref{figM}.

With the help of Eq.~(\ref{Eqc}),
we can derive the energy dispersion of the perfect surface state
with the derived current expression in Eq.~(\ref{EqI2}).
In fact, we are able to arrive at an identical expression as is given in Eq.~(\ref{EqEe}).
This consistency indicates the reliability of our analytical derivation.

According to Eq.~(\ref{Eq170304_03}), the magnetic moment and current are related by $M_{z}=\pi R_{L}^{2}I_{e}$.
With the current expression in Eq.(\ref{EqI}) substituted and $\langle s_{z}\rangle$ neglected,
we have
\begin{equation}
\frac{M_{z}}{e/ h}=\pi R_{L}|A|-\frac{\pi}{2}m_{0} .
\label{Eq170304_02}
\end{equation}
The second term is much smaller than the first term in our parameter regime and then it explains the linear dependence of the magnetic moment on QD radius.
Eqs.~(\ref{EqI2}) and (\ref{Eq170304_02}) combine to lead to $M_zI_e = \frac{\pi e^2}{h^2} (|A|-\frac{m_0}{2R_L})^2$, it reduces into Eq.~(\ref{EqIM}) straightforwardly with the higher order term neglected.

\subsubsection{Spin orientation}\label{sec_spin}

As is well known that the spin and momentum will be locked for the edge or surface state of
topological materials.~\cite{Hasan2010RMP,Qi2011RMP}
Recently, the spin texture of Fermi arc in Weyl semimetal is also observed directly
by ARPES measurements.~\cite{Lv2015PRL,Das2016ncomm}
Here we show that there is also a corresponding relation in our Weyl semimetal QD system.
Considering that momentum is not a good quantum number now in the absence of translation invariance, we replace momentum with current and study the relation between spin orientation and the flow direction of current.

The density distribution of spin can be calculated
by $\langle s_i\rangle = \frac{1}{2}\Psi^\dag \sigma_i\Psi $.
For the surface state, the radial ($\langle s_r\rangle$), tangential ($\langle s_\varphi \rangle$)
and axial ($\langle s_z\rangle$) components can be derived as
\begin{align}
&  \langle s_r\rangle = \frac{2{\rm sin}^2(n_z\pi z/L)}{L} |A_{m-1}(r)B_m(r)|
\mathrm{cos} \delta\theta,\nonumber\\
&  \langle s_\varphi \rangle = \frac{2{\rm sin}^2(n_z\pi z/L)}{L}|A_{m-1}(r)B_m(r)|
\mathrm{sin} \delta\theta,\nonumber\\
&  \langle s_z\rangle =\frac{2{\rm sin}^2(n_z\pi z/L)}{L}(|A_{m-1}(r)|^{2} - |B_m(r)|^{2})/2,
\end{align}
where $\delta\theta = \theta_{B}-\theta_{A}$ determines the phase difference
between the radial wave function $A_{m-1}(r)$ and $B_m(r)$.
Fortunately, $\delta\theta$ is a constant and can be taken to be $-\pi/2$ according to our numerical calculation and discussions above. It vanishes the radial component, $ \langle s_r\rangle = 0$,
and indicates that the spin lies in the plane of surface state.~\cite{Lee2009PRL,Kundu2011PRB}
The projection of spin in the horizontal plane is antiparallel to the azimuthal direction, which suggests a clockwise distribution.
As we have shown, it is also clockwise for the probability current density, $\vec{J}=-|J_\varphi| \vec{e}_\varphi $.
Therefore the projection of spin on the horizontal plane will be parrel with the current.
Our numerical calculation suggests that $\langle s_z\rangle$ is small enough to be neglected (see Fig.\ref{figs} at $B=0$).
Then we arrive at
\begin{align}\label{Eqhelicity}
\frac{\vec{J} \cdot\vec{s} }{|\vec{J} | |\vec{s} |} \approx 1 .
\end{align}
It is similar with the definition of helicity which is the projection of spin on the moving direction. Eq.~(\ref{Eqhelicity}) suggests that an electron in the surface state has a positive helicity or chirality.
In fact, the value of $\frac{\vec{J} \cdot\vec{s} }{|\vec{J} | |\vec{s} |}$ is mainly determined by the topological properties of the system. It corresponds to the spin-momentum locking in 3D case.

\begin{figure}[ptb]
\centering
\includegraphics[width=1\columnwidth]{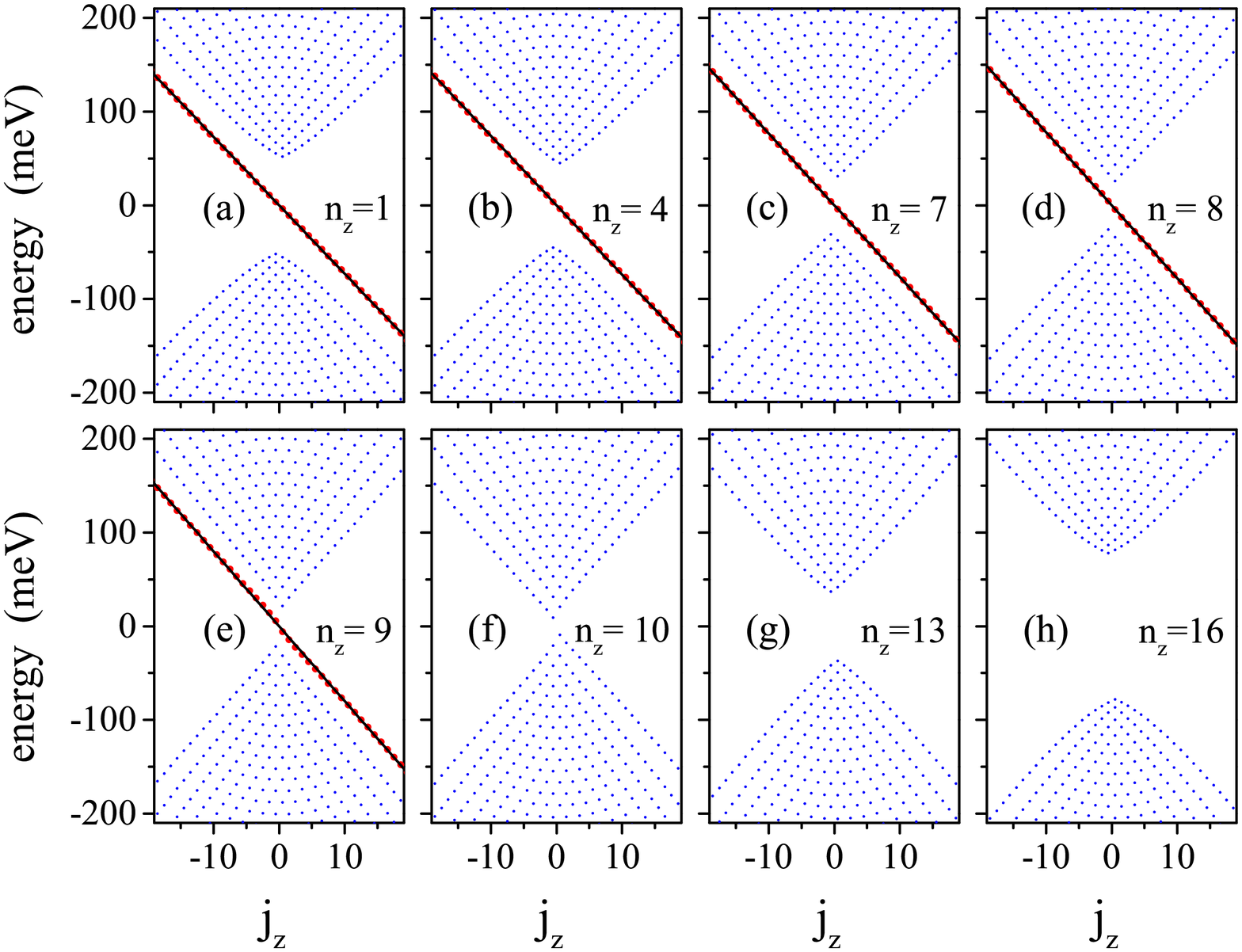}\newline
\caption{
(Color online) The energy spectra corresponding to $n_z=1$, $4$, $7$, $8$, $9$, $10$, $13$, and $16$.
Red dots in panels with $n_z=1$ to $9$ show the linear band
while the black line is the fitting line by Eq.~(\ref{EqEe}).}
\label{fig_nz_01}%
\end{figure}

\subsection{Quantization in the axial direction}

In this subsection we reveal the effect of quantization in the axial direction
which is denoted by $n_z$. We construct the relation between properties of Weyl semimetal QD
and Weyl semimetal. Besides, we also study the size effect.

In Fig.~\ref{fig_nz_01}, we plot the energy spectra for $n_z=1$, $4$, $7$, $8$, $9$, $10$, $13$, and $16$.
We find that there is a linear band in the gap regime for $n_z=1$, $4$, $7$, and $8$,
which is fitted well by the black line due to Eq.~(\ref{EqEe}).
The linear band disappears for $n_z=10$, $13$, $16$ leaving a bare energy gap.
The case of $n_z=9$ lies in the crossover regime, though we also use Eq.~(\ref{EqEe}) to fit the band.
The direct energy gap decreases gradually with rasing $n_z$ at first for $n_z\leq10$,
reaches the minimum at $n_z=10$ and then increases with raising $n_z$ any more.
It indicates that the linear band, which has been proved to be the band of surface states
in the discussion above, approaches the bulk band gradually as the direct gap decreases
and merges into the bulk band at the minimum of the bulk gap.

\begin{figure}[ptb]
\centering
\includegraphics[width=1\columnwidth]{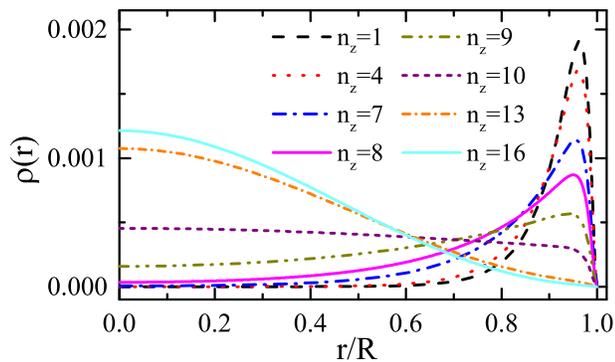}\newline
\caption{
(Color online) Radial electron density distributions of the eigenvectors
with $j_z=1/2$, $n_r=0$ but several $n_z$.}%
\label{fig_nz_02}%
\end{figure}

The linear band for small $n_z$ ($n_z\leq8$) is expected naturally to be surface states as is shown in Fig.~\ref{fig_nz_02} where we plot and study the evolution of the radial electron density distributions for eigestates with $j_z=1/2$, $n_r=0$ and several $n_z$.
The $n_r=0$ band corresponds to the linear band for $n_z<9$ while the first conduction band for $n_z\ge 9$, respectively.
We have a surface state for $n_z=1$, $4$, $7$, $8$, while the weight of distribution shifts from the side surface to the bulk of the QD with increasing $n_z$.
The density distribution of $n_z=9$ lies in the crossover regime with a considerable weight both around the side surface and spreading in the bulk. For $n_z=10$, $13$ and $16$, the weight has shifted mostly into the bulk.
It indicates that the eigenstate denoted by $j_z=1/2$ and $n_r=0$ evolves from the original surface state into the bulk state gradually with increasing $n_z$.

Our numerical results indicate that there will be both surface and bulk states near the Fermi level
due to sub-bands corresponding to different $n_z$,
which makes the Weyl semimetal QD distinguished from the topological insulator QD ~\cite{Chang2011PRL} and conventional semiconductor QD.
It is consistent with the coexistence of Weyl point (bulk states) and Fermi arc (surface states) in 3D Weyl semimetal and may be taken as its projection in zero-dimensional QD system.
The minimum of the bulk energy gap corresponds to the Weyl point which is the touch point of valence and conduction bands.
The chirality of Weyl points determines the flow direction of the vortex current of surface states.
Surface states of the Weyl semimetal QD emerge for several $n_z$ only,
which is the correspondence that surface states of 3D Weyl semimetal exist
only for several momentum to form the Fermi arc.
It is fortunate that these correspondences exist and it allows the study of Weyl semimetal in the QD system.

\begin{figure}[ptb]
\centering
\includegraphics[width=1\columnwidth]{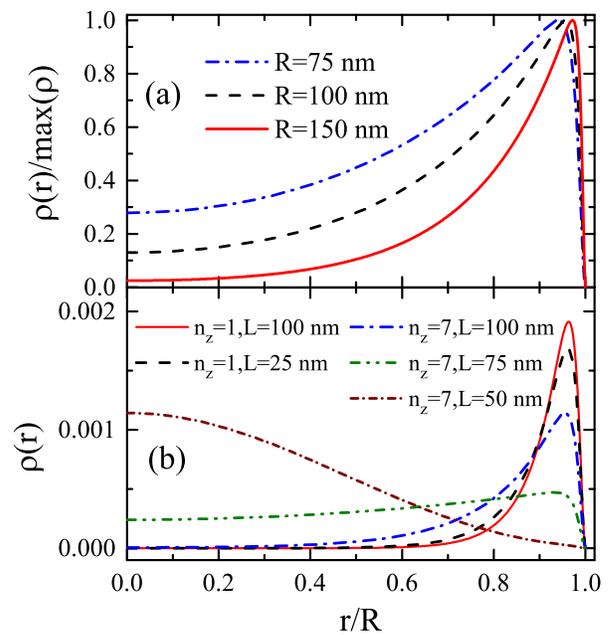}\newline
\caption{(Color online) Radial electron density distribution $\rho(r)$ as a function of QD radius $R$ (a)
and QD height $L$ (b). The eigenvector is taken to be $n_z=9$, $j_z=1/2$, $n_r=0$ (a) and $n_z=1$ and $7$, $j_z=1/2$, $n_r=0$ (b).}%
\label{fig_nz_03}%
\end{figure}

The location of the minimum of direct bulk gap borders on the region with surface states and it is determined by the spacing of two Weyl points of 3D Weyl semimetal.
If we take $k_z =n_z\pi/L$ as a constant, then the Hamiltonian in Eq.~(\ref{EqHc})
becomes a two-dimensional system, with $M_{n_z}$ acting as the mass term.
A negative $M_{n_z}$ leads to a normal insulator phase, while a positive one leads to a quantum anomalous Hall phase with edge states which are just the surface states in our QD system.
The criterion $M_{n_z}>0$ demands $k_z=n_z\pi/L <k_0$ for the emergence of surface states, where $k_0$ is half of the spacing between two Weyl points.
It explains why surface states emerge for small $n_z$ while disappear for large $n_z$.

This criterion predicts that there will be surface states in the regime $n_z=1\sim 10$ for parameters we have chosen.
However, our numerical results show that the $n_z=9$ case lies in the crossover regime and there is no surface state for $n_z=10$ case.
This inconsistency is due to the size effect.
To show this point, we plot the radial electron density distribution $\rho(r)$ of the eigenstate with $j_z=1/2$, $n_r=0$, $n_z=9$ for several QD radius $R$.
The result is shown in Fig.~\ref{fig_nz_03}(a), and it's clear that the weight shifts
from the bulk to the side surface as we increase the QD radius.
Once the radius is large enough it turns out to be a surface state.
Therefore, it is the size effect that breaks the surface state for $n_z$ neighbouring the critical value
and results in the crossover regime.

Fig.~\ref{fig_nz_03}(b) plots the radial electron density distribution $\rho(r)$
to show the effect of QD height $L$.
It's clear that $\rho(r)$ locates around the side surface for a large $L$
for the chosen two eigenstates.
Decreasing $L$ will shift the weight to the bulk.
The $n_z=7$ eigenstate becomes a bulk state already for $L=50$ nm
while it is still a surface state even for $L=25$ nm in the case of $n_z=1$ eigenstate.
It indicates that a large QD height is in favor of the emergence of surface states
and the eigenstates with large $n_z$ will be effected more significantly by QD height,
which can be inferred from the criterion $M_{n_z} = M - \frac{1}{2}m_{0} (\frac{n_z\pi}{L})^2>0$.
An extreme case is that if $L$ is small enough to make $M_{n_z}<0$ for any $n_z$,
then there will be no surface states any more.
This effect provides a feasible method to control the surface state.

\section{The effect of magnetic field on surface states}

In this section we study the property of surface states in the presence of a uniform magnetic field $\mathbf{B}=(0,0,B)$ along axial direction. By studying the evolutions of energy levels and radial electron density distribution, we show that magnetic fields will break topological surface states gradually to form Landau levels. The location of the crossover regime separating these two phases is determined with a approximate derivation. Besides, we also study the spin polarization. Without loss of generality, we focus on the $n_z=1$ case to show these interesting properties.

Magnetic fields will induce two types of contributions, the orbital and Zeeman effects.
The Zeeman term takes the same form as $\tilde \Delta_z\sigma_z$ in the Hamiltonian given in Eq.~({\ref{EqHc}),
but with a smaller magnitude.
Therefore it is neglected in the following discussion.
The orbital effect can be included by a Peierls substitution $\mathbf{k}\rightarrow\mathbf{\pi}=\mathbf{k}+\frac{e}{\hbar}\mathbf{A}$, where $\mathbf{A}$ is the vector potential and $e$ is the magnitude of the elementary charge. Then the Hamiltonian in Eq.~({\ref{EqHc}) becomes %
\begin{equation}\label{Eq170416_01}
H=\left(
\begin{array}
[c]{cc}%
M-\frac{m_{0}}{2}(\pi_{x}^{2}+\pi_{y}^{2} +k_z^2) & A\pi_{-}\\
A\pi_{+} & -[M-\frac{m_{0}}{2}(\pi_{x}^{2}+\pi_{y}^{2} +k_z^2)]
\end{array}
\right),
\end{equation}
in which $\pi_\pm = \pi_x \pm \pi_y$.
For a cylinder geometry, we take the gauge as
$\mathbf{A}=\mathbf{B}\times\vec{r}/2$, which becomes $\mathbf{A} =\frac{1}{2}Br\vec{e}_{\theta}$
in the cylindrical coordinate system.
It is clear that the rotation invariance survives and
there is no $z$- and $\varphi$-dependent terms in the vector potential.
Therefore, the two variables $z$ and $\varphi$ can be separated the same as
we have done in the absence of magnetic field,
making $n_z$ and $j_z$ still good quantum numbers.
And the eigenvector shares the identical form as is given in Eqs.~(\ref{Eq170118_01}) and (\ref{Eq170117_01}),
then it's enough to modify the radial Hamiltonian by adding an extra term
\begin{eqnarray}\label{EqHB}
\delta H=\frac{m_0}{2} \tilde B\sigma_0 + A \tilde B r\sigma_y - m_0 \tilde B j_z \sigma_z - \frac{m_0}{2} \tilde B^2 r^2\sigma_z,
\end{eqnarray}
in which $\sigma_0$ is a $2\times2$ identity matrix and $\tilde B=\frac{eB}{2\hbar}$ proportional to
the magnetic field with magnitude $7.583\times10^{-4}/({\rm nm})^{2}$ for $B =1$T.
The first term is a trivial constant to shift the energy spectra as a whole.
The second and last two terms are to revise the spin-orbit coupling and Zeeman terms respectively.

\begin{figure}
  \centering
  \includegraphics[width=1\columnwidth]{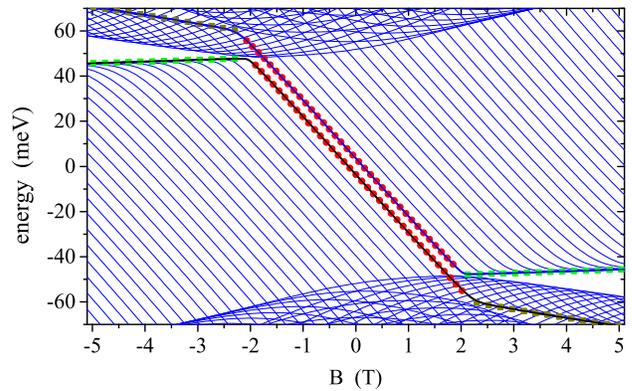}\\
  \caption{(Color online) The energy levels as a function of the magnetic field in the $n_z=1$ case.
  Red circles are the fitting according to Eq.~(\ref{EqB})
  for the eigenstates of $j_z=\frac{1}{2}$ (the black line) and $j_z=-\frac{1}{2}$ (the blue line) with fixed $n_r=0$. Light green and dark yellow squares fit the zeroth and $N=1$ modes of Landau levels respectively according to Eq.~(\ref{EqLandau}).  }\label{figm}
\end{figure}

The evolution of energy levels against the magnetic field is plotted in Fig.~\ref{figm}.
A detailed analysis indicates that the energy spectra displays a symmetry
\begin{eqnarray}\label{Eq170413_02}
E_{-j_{z}}(-B)=-E_{j_{z}}(B).
\end{eqnarray}
In absence of magnetic field it reduces to Eq.~(\ref{Eq170413_01}), which suggests that this relation is also
due to the antiunitary transformation $P$ given in Eq.~(\ref{EqP}).
It's straightforward to prove that $P H_{-j_{z},-B}P^{\dag} = - H_{j_{z},B}$ which leads to
the symmetry of energy spectra given in Eq.~(\ref{Eq170413_02}) and a constrain on corresponding eigenvectors
\begin{align}\label{Eq170419_01}
\psi_{j_{z},B} =  P \psi_{-j_{z},-B}.
\end{align}
\begin{figure}
  \centering
  \includegraphics[width=1\columnwidth]{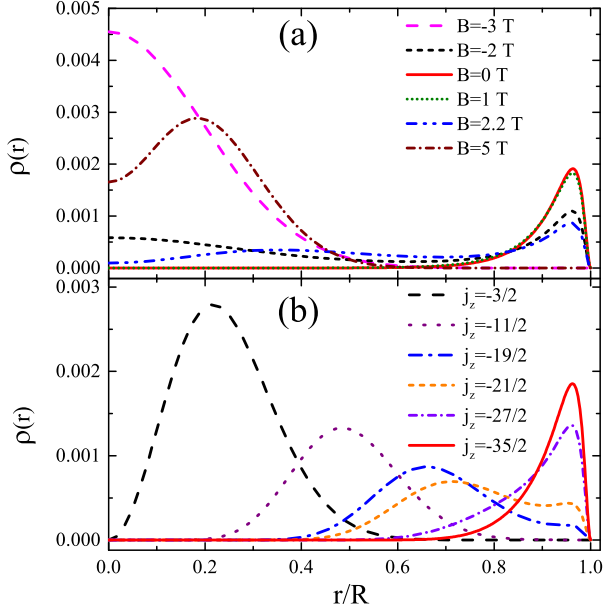}\\
  \caption{(Color online) Radial density distributions $\rho(r)$ of eigenstates with
  with $n_z=1$ and $n_r=0$ for several magnetic fields at fixed $j_z=\frac{1}{2}$ (a)
  and for several angular momentums at fixed $B=5$ T (b).}\label{figmw01}
\end{figure}
The energy level of the specific eigenstate with $j_z=\frac{1}{2}$ and $n_r=0$ is plotted as the black line in Fig.~\ref{figm}. It can be divided into three linear regimes with different slopes.
The lines corresponding to eigenstates of $n_r=0$ but with different $j_z$ are nearly parallel in the weak field regime
but degenerate in both positive and negative strong field regimes.
There is a crossover regime separating the weak and strong regimes and its location $B_c$ varies between eigenstates labeled with $j_z$.

To further understand the weak and strong field regimes,
we plot the radial electron density distributions of eigenstates for several magnetic fields
with fixed angular momentum $j_z=\frac{1}{2}$ in Fig.~\ref{figmw01}(a)
and for different $j_z$ with fixed magnetic field $B=5$ T in Fig.~\ref{figmw01}(b).
It's clear that the eigenvector is a surface state in the weak field regime for curves with $B=0$, $1$ T
in Fig.~\ref{figmw01}(a) and $j_z=-\frac{35}{2}$, $-\frac{27}{2}$ in Fig.~\ref{figmw01}(b).
However, eigenvectors distribute in the bulk of the QD in the strong field regime for curves with $B=-3$T
and $5$T in Fig.~\ref{figmw01}(a) and $j_z=-\frac{3}{2}$ and $-\frac{11}{2}$ in Fig.~\ref{figmw01}(b).
Eigenstates with $B=-2$T and $2.2$T in Fig.~\ref{figmw01}(a) and $j_z=-\frac{19}{2}$ and $-\frac{21}{2}$
in Fig.~\ref{figmw01}(b) lie in the crossover regime,
and we find that the electron density distribution is a mixing of states in the bulk and on the side surface.

Figure \ref{figm01} displays the energy spectra for several magnetic fields $B=0$, $1$, $2.2$, and $\pm5$ T.
The $n_r=0$ band is linear in Fig.~\ref{figm01}(b,c) with $B=0$ and $1$T lying in the weak magnetic field regime.
As studied above, the linear band in the gap regime corresponds to surface states.
It moves downward with rasing magnetic field, partly merges with the bulk bands for $B=2.2$ T.
It becomes complex for $B =5$ T.
There is a linear band for $j_z<-\frac{17}{2}$ but a flat band with high degeneracy for $-\frac{17}{2}<j_z < 0$,
lying in the weak and strong field regimes respectively, concluded from Fig.~\ref{figm}.
The case of $B=-5$T is similar with the $B=5T$ case, with eigenenergy and $j_z$ reversed
due to the $P$ symmetry given in Eq.~(\ref{Eq170413_02}).
Besides, we also notice that there is a dip and hump structure connected to the flat energy band in Fig.~\ref{figm01} (a) and (e) respectively, which is similar to the structure predicted in the nanoribbon of quantum spin Hall~\cite{Chen2012PRB} and quantum anomalous Hall insulators.~\cite{Zhang2014PRB}
It suggests that the dip and hump structures may be universal in topological nontrivial materials
in presence of strong magnetic field.

\begin{figure}
  \centering
  \includegraphics[width=1\columnwidth]{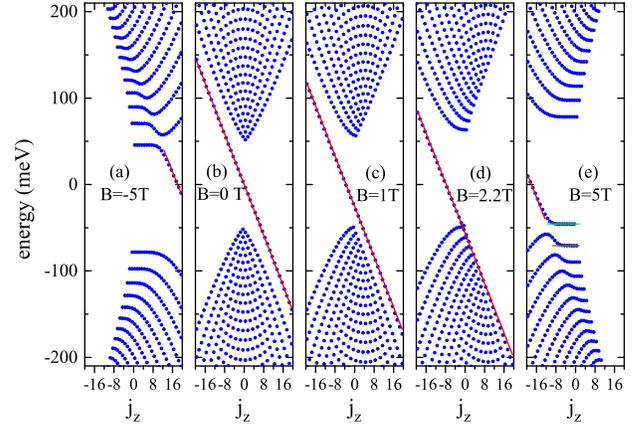}\\
  \caption{(Color online) Energy spectra for magnetic fields $B=-5$, $0$, $1$, $2.2$, and $5$T
  in the case of $n_z=1$. Red lines are fitting lines according to Eq.~(\ref{EqB}).
  The light green and dark yellow lines in panel (e) fit the zeroth and $N=1$ mode of Landau levels, respectively.}\label{figm01}
\end{figure}

By assuming a perfect surface state the same as that in Eq.~(\ref{Eq170523_01}),
we can derive a simple expression of the energy spectra of the surface band,
\begin{equation}\label{EqB}
E_{j_z}(B)=-j_{z}(\frac{|A|}{R_{L}}-\frac{m_{0}}{2R_{L}^{2}}) - |A|\tilde BR_{L} +\frac{1}{2}m_{0}\tilde B,
\end{equation}
which reduces to Eq.~(\ref{EqEe}) in the absence of magnetic field.
It depends linearly on magnetic field and angular momentum, consistent with numerical results
in the weak magnetic field regime.
The linear bands in the weak field regime are fitted well with this expression as red circles in Fig.~\ref{figm} and red lines in Fig.~\ref{figm01}, respectively.
In the strong field regime, magnetic field makes the weight of distribution shifts from the surface to the bulk to break the surface state as is shown in Fig.~\ref{figmw01}(a) for eigenstate with $j_z=\frac{1}{2}$.
Resultantly, Eq.~(\ref{EqB}) lacks the high degeneracy of eigenstates with different $j_z$ and fails to describe the degenerate band in the strong field regime.
The critical magnetic field depends on $j_z$. Therefore, there will be states with higher $j_z$ still distributing near the surface for a large magnetic filed, which originate from both the nontrivial topology and magnetic field, as is shown by the states fitted with red lines in Fig.~\ref{figm01}(a) and (e).

In general, a strong magnetic field will lead to Landau quantization.\cite{addLH}
Therefore it's also expected that Landau quantization occurs in the Weyl semimetal QD system
in the strong magnetic field regime.
Landau levels of a infinite system can be derived to be
\begin{eqnarray}\label{EqLandau}
E_{N}^{(e)} & =& m_0 \tilde B + \sqrt{ (M_{n_z} - 2m_0 |\tilde B| N)^2 + 4A^2 |\tilde B| N },\nonumber\\
E_0& = &m_0 \tilde B -sign(\tilde B) M_{n_z},   \nonumber\\
E_{N}^{(h)} &=& m_0 \tilde B - \sqrt{ (M_{n_z} - 2m_0 |\tilde B| N)^2 + 4A^2 |\tilde B| N }, \nonumber\\
\end{eqnarray}
in which $N=1$, $2$, $...$, $E_0$ is the zeroth mode, and $E^{(e)}$ ($E^{(h)}$) describes the electron-like (hole-like) Landau levels.
Degenerate magnetic levels in Fig.~\ref{figm} and flat bands in Fig.~\ref{figm01}(e)
are well fitted by the zeroth and $N=1$ Landau levels,
which indicates the well formation of Landau levels in the strong magnetic field regime.
The eigenvectors belonging to the same Landau level are expected to share the similar form
but located in different positions in the bulk of QD as shown in Fig.~\ref{figmw01}(b).
Based on these results, it concludes that the strong filed regime lies in the Landau quantization phase.

\begin{figure}
  \centering
  \includegraphics[width=1\columnwidth]{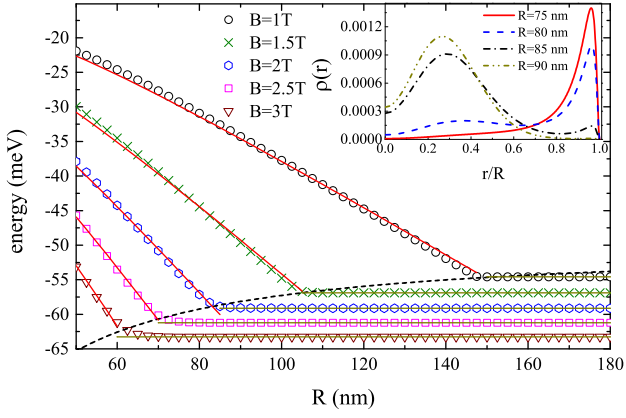}\\
  \caption{(Color online) Energy levels vs QD radius $R$ for magnetic fields $B=1$, $1.5$, $2$, $2.5$, and $3$ T.
  Red and dark yellow lines are fitting by Eqs.~(\ref{EqB}) and (\ref{EqLandau}) respectively.
  The black dash line is to fit the location of crossover regime.
  The inset shows radial density distributions for several QD radius $R$ at fixed magnetic field $B=2T$.
  Other parameters are $n_z=1$, $j_z=\frac{1}{2}$, and $n_r=0$.}\label{figmR}
\end{figure}

Combining Eqs.~(\ref{EqB}) and (\ref{EqLandau}), it's able to determine the location of the crossover regime between the weak and strong field phases. The transition from the surface state to the zeroth mode occurs at
\begin{eqnarray}\label{Eq170416_02}
   \tilde B_c = \frac{-j_z (\frac{A}{R_L} - \frac{m_0}{2R_L^2} ) + sign(\tilde B) M_{n_z}} {AR_L + \frac{m_0}{2}},
\end{eqnarray}
which also determines the beginning of Landau quantization.
The analytical expression of the critical magnetic field $\tilde B_c$ transiting from the surface state
to the $N=1$ Landau level can also be derived by setting $E_1^{(e/h)} =E_{j_z}(B)$, in which $E_{j_z}(B)$ is the dispersion of surface states given in Eq.~(\ref{EqB}).
But the expression of $\tilde B_c$ is in a tediously long form which is not given here.
The calculated values lie approximately in the crossover regime in Figs.~\ref{figm} and \ref{figm01}.

\begin{figure}
  \centering
  \includegraphics[width=1\columnwidth]{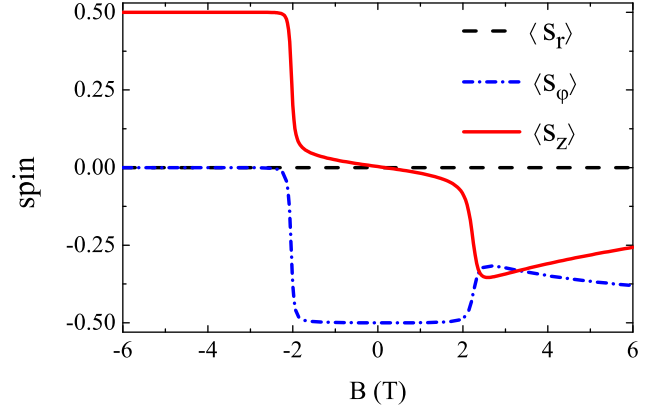}\\
  \caption{(Color online) Spin polarization vs magnetic field for the eigenstate $j_z=\frac{1}{2}$, $n_r=0$, $n_z=1$, the black line in Fig.~\ref{figm}.}\label{figs}
\end{figure}

In Fig.~\ref{figmR}, we plot the energy level as a function of QD radius $R$ for magnetic fields $B=1$, $1.5$, $2$, $2.5$, and $3$ T.
The energy level decreases with raising $R$ at first and then saturates for $R$ larger than a critical value $l_B$.
It's as expected that $l_B$ will decrease with raising magnetic field.
In the inset, we plot the radial electron density distribution $\rho(r)$
for $R=75$, $80$, $85$, and $90$ nm at fixed $B=2$ T.
We find that it is a surface state for $R=75$ nm, an eigenvector of Landau level for $R=90$ nm similar with that in Fig.~\ref{figmw01}, and a mixing of those two states for $R=80$ and $85$ nm.
The curves of the levels versus the magnetic field
before and with the saturation value can be fitted by the red and dark yellow lines
due to Eq.~(\ref{EqB}) and Landau level $E_{1}^{(h)}$, respectively.
The position of the crossover regime, $l_B$, is well fitted by the black dash line derived by combining Eq.~(\ref{EqB}) and Landau level $E_{1}^{(h)}$.
It indicates that Landau quantization phase occurs for large enough QD radius,
while the phase with only surface states appear at a moderate regime of QD radius.
Here we point out that a more precise magnetic length $l_B$ and criterion $R>l_B$
is given instead of the usual one \cite{Chang2011PRL} $R\gg l_B = \frac{\hbar}{|eB|}$
to judge whether the magnetic field and QD size are large enough to realize Landau quantization.
Besides, we notice that increasing magnetic field makes the level depend on $R$ more linearly before saturation.
It is consistent with Eq.~(\ref{EqB}) since the second term of Eq.~(\ref{EqB}) is proportional to magnetic filed and will be dominate for a large magnetic field.

In Fig.~\ref{figs}, we plot the spin polarization of the eigenstate $j_z=\frac{1}{2}$ and $n_r=0$, the one denoted as a black line in Fig.~\ref{figm}.
The radial component $\langle s_r \rangle$ vanishes exactly since the unitary matrix given in Eq.~(\ref{EqS})
makes the Hamiltonian in Eq.~(\ref{Eq170416_01}) real to lead to the phase difference $\delta \theta = -\pi/2$, as is discussed in absence of magnetic field.
The angular and axial components $\langle s_\varphi \rangle$ and $\langle s_z \rangle$
can be divided into three regions
which correspondes to Landau level $E_0$, the nontrivial surface state, Landau level $E_1^{(h)}$, successively from left to right.
In the regime of zeroth Landau level $E_0$, spin is polarized in the axial direction with $\langle s_z\rangle=\frac{1}{2}$ and the other components vanished.
In the surface state regime, we have $\langle s_\varphi \rangle \approx\frac{1}{2}$
while $s_z$ is small, suggesting that spin aligns antiparallel to the tangential direction approximately.
And the spin density distributes around the side surface in this regime.
In the $E_1^{(h)}$ regime, $s_z$ ($s_\varphi$) increases (decreases) with raising magnetic field
and spin is not locked into a specific direction any more.

\section{Discussion and Conclusion}

We have proposed and studied the Weyl semimetal QD.
Our results reveal that there will be both surface and bulk states coexisting near the Fermi level.
We focus on properties of the surface state,
which locates only on the side surface of the cylinder QD studied here.
Its energy dispersion depends linearly on total angular momentum leading to a chiral current and
orbital magnetic moment with their product being a constant.
The spin orients clockwisely in the horizontal plane near the side surface.
A strong magnetic field along the axial direction will destroy the surface state
and result in the Landau quantization.
The spin will deviate away from the horizontal plane and become completely polarized
in the axial direction for the zeroth Landau level.

Our results show the projection of properties of 3D Weyl semimetal in the zero-dimensional QD system.
Surface states will emerge only for some certain quantum numbers determined by the spacing of Weyl points,
which act as the correspondence of the Fermi arc.
There is a minimum of direct energy gap with its location related to the projection of the Weyl point.
It provides the possibility of study the Weyl physics in the Weyl semimetal QD system
which may be fabricated and tuned more easily.
Due to the large surface-to-volume ratio, the contribution of surface-state carriers
are amplified and it suggests a new method to verify the topological nontrivial property of Weyl semimetal by separating the signal of the surface state, which may be realized in quantum transport measurements.

In this paper, we study the Weyl semimetal QD with a low-energy effective Hamiltonian
with only two Weyl points.
As is well known, the case of real materials may be much more complex since there will be several pairs of Weyl points which may locate away from the Fermi surface.~\cite{Burkov2016nmater,Jia2016nmater}
However, this simple model allows us to pick out the key characteristic and makes our results accessible
for plenty of materials.
Parameters in the Hamiltonian may vary with the system size, resulting in a different dependence on QD size.
They may also be tuned by external factors to realize more exotic properties and applications.
A further understanding needs more theoretical and experimental studies.
Another effect missed here of locating the electron in a QD is the emergence of Coloumb interaction.
Due to the coexistence of surface and bulk states, it's expected that there will be interactions not only among electrons in the bulk states but also among electrons in the surface states and between electrons occupying surface and bulk states. These new terms will give rise to new phenomena absent
in the topological insulator QDs and conventional semiconductor QDs.
In a few words, the further study of Weyl semimetal QD will pave a new way for the investigation of Weyl semimetal.

\section*{ACKNOWLEDGEMENTS}
This work was supported by the National Basic Research Program of China (Grants No.2017YFA0303301, No. 2015CB921102), the National Natural Science Foundation of China (Grants No.11274364, No.11574007) and the Doctoral Foundation of University of Jinan (Grant No. XBS160100147).


\begin{thebibliography}{80}
\bibitem{note01}
Be careful that the dimension of angular momentum is taken to be 1 in this work. Therefore the right side of Eq.~(\ref{Eqc}) has the same dimension as electric current.
\end{thebibliography}
\end{document}